\documentclass{PoS}

\usepackage{wrapfig}
\usepackage{graphicx}% Include figure files

\def\figdir{./}

\newcommand\figwidth{.48\textwidth}
\newcommand\Eq[1]{Eq.~\ref{eq:#1}}
\newcommand\Fig[1]{Fig.~\ref{fig:#1}}

\title{Photon mass term as an IR regularization for QCD+QED on the lattice}

\ShortTitle{Photon mass term as an IR regularization for QCD+QED on the lattice}

\author{%
\speaker{Michael G. Endres} \\ %
Center for Theoretical Physics, Massachusetts Institute of Technology, Cambridge, MA 02139, USA\\
E-mail: \email{endres@mit.edu}
}

\author{%
Andrea Shindler \\
IAS, IKP and JCHP, Forschungszentrum J\"{u}lich, 52428 J\"{u}lich, Germany \\
E-mail: \email{a.shindler@fz-juelich.de}
}

\author{%
Brian C. Tiburzi \\
Department of Physics, The City College of New York, New York, NY 10031, USA \\
Graduate School and University Center, The City University of New York, New York, NY 10016, USA \\
RIKEN BNL Research Center, Brookhaven National Laboratory, Upton, NY 11973, USA \\
E-mail: \email{btiburzi@ccny.cuny.edu}
}

\author{%
Andr\'{e} Walker-Loud \\
Nuclear Science Division, Lawrence Berkeley National Laboratory, Berkeley, CA 94720, USA \\
Department of Physics, College of William and Mary, Williamsburg, VA 23187-8795, USA \\
Jefferson Laboratory, 12000 Jefferson Avenue, Newport News, VA 23606, USA \\
E-mail: \email{awalker-loud@lbl.gov}
}

\abstract{%
Inclusion of QED in lattice QCD calculations can lead to power-law volume artifacts as a consequence of the long-range nature of the interaction.
Such artifacts must be removed by extrapolation in order to attain reliable infinite volume estimates of observables and quantities derived from them.
As an alternative to this methodology, we  consider the use of a photon mass term as an infrared regulator for QCD+QED, and explore the viability of its use in determining hadron mass shifts and splittings. 
}

\FullConference{%
The 33rd International Symposium on Lattice Field Theory\\
14 -18 July 2015\\
Kobe International Conference Center, Kobe, Japan}

\begin{document}

Sources of isospin-breaking in nature include strong breaking due to the quark mass differences, which arise from differences in Yukawa couplings in the weak sector of the Standard Model, and electromagnetic breaking due to the charge differences between quarks. 
Understanding the delicate interplay between these effects from first-principles requires a nonperturbative treatments of study, for which lattice techniques are ideally suited.
A number of strategies for introducing QED interactions in lattice QCD have been developed in recent years \cite{Duncan:1996xy, Duncan:1996sq, Lehner:2015bga, Lucini:2015hfa}, and an analytic understanding of the finite volume effects achieved with the use of effective field theories \cite{Davoudi:2014qua, Fodor:2015pna}.
The numerical work which utilizes these developments, however, often demand large volumes in order to gain adequate control over systematic uncertainties in the requisite power-law infinite volume extrapolations of observables.
In this study, we explore the use of a photon mass as an alternative means for handling IR artifacts.
To test the viability of this new approach, we compare determinations of the hadronic mass shifts and splittings due to electromagnetic effects in electroquenched QCD, obtained in two ways:
\begin{enumerate}
\item A conventional infinite volume extrapolation of estimates, where electromagnetic interactions are included using a Coulomb gauge-fixed zero-mode subtracted photon action \cite{Portelli:2010yn};
\item A vanishing photon mass extrapolation of infinite volume estimates, where electromagnetic interactions are included via a massive photon action.
\end{enumerate}
Since this work is already described in depth elsewhere \cite{Endres:2015gda}, we take this opportunity to summarize some of the main results, and expand upon several aspects of the study.

For the comparisons described above, we consider electroquenched QCD+QED on a Euclidean space-time lattice with lattice spacing $a$, spatial extent $L_j = L$ ($j=1,2,3$), and temporal extent $L_0 = T$.
Although we introduce a gauge-symmetry violating photon mass term in the latter case, prior gauge-fixing of the QED action is helpful as it enables reliable stochastic estimates of charged correlators in the limit of small photon mass.\footnote{Without such gauge fixing, charged correlator ratios suffer from a signal/noise problem in the limit $m_\gamma/m_\pi \to 0$.}
For the massive QED study, we consider a $R_\xi$ gauge-fixed non-compact QED action, noting that this choice of gauge fixing is the only one that retains full hypercubic symmetry of the lattice at finite photon mass, $m_\gamma$.
In momentum space,  the action for the vector field $\tilde A_\mu(p)$ is given by
\begin{eqnarray}
S_\gamma = \frac{1}{8\pi \alpha}\frac{1}{L^3 T}\sum_p \sum_{\mu\nu} \tilde A_\mu(p) \tilde G_{\mu\nu}(p) \tilde A_\nu(p)\ ,
\label{eq:massive_qed_action}
\end{eqnarray}
where $G_{\mu\nu}(p) = (p^2 +m_\gamma^2) \delta_{\mu\nu} + (1/\xi -1)p_\mu p_\nu$ and $p_\mu = (2/a) \sin(2\pi n_\mu a/L_\mu)$ for integer $n_\mu \in [0, L_\mu/a)$, and $\alpha = e^2/(4\pi) \approx 1/137$.
Note that $G_{\mu\nu}(p)$ has eigenvalues $p^2/\xi+m_\gamma^2$ and $p^2 +m_\gamma^2$, with the latter being three-fold degenerate for $p\neq 0$.
In the limit $\xi\to 0$, corresponding to Landau gauge, the mode proportional to $p_\mu$ decouples from the theory.
In contrast with typical treatments of QED on the lattice, we retain the (massive) zero-modes rather than explicitly removing them from the theory.
In this formulation of QCD+QED, mass shifts and mass splittings receive $\textrm{O}(\alpha)$ corrections from three sources: the presence of \textit{zero modes}, the presence of a \textit{finite volume}, and the presence of a  \textit{finite photon mass}.
All three effects can be accounted for systematically via an effective field theory (EFT) description for hadrons of mass $M$ (e.g., $m_\pi$, $m_K$, $m_n$ and $m_p$) and charge $Q$, expanding perturbatively in the Compton wavelength of the hadron.

\begin{wrapfigure}{R}{0.5\textwidth}
%\begin{figure}
\centering
\includegraphics[width=\figwidth]{\figdir 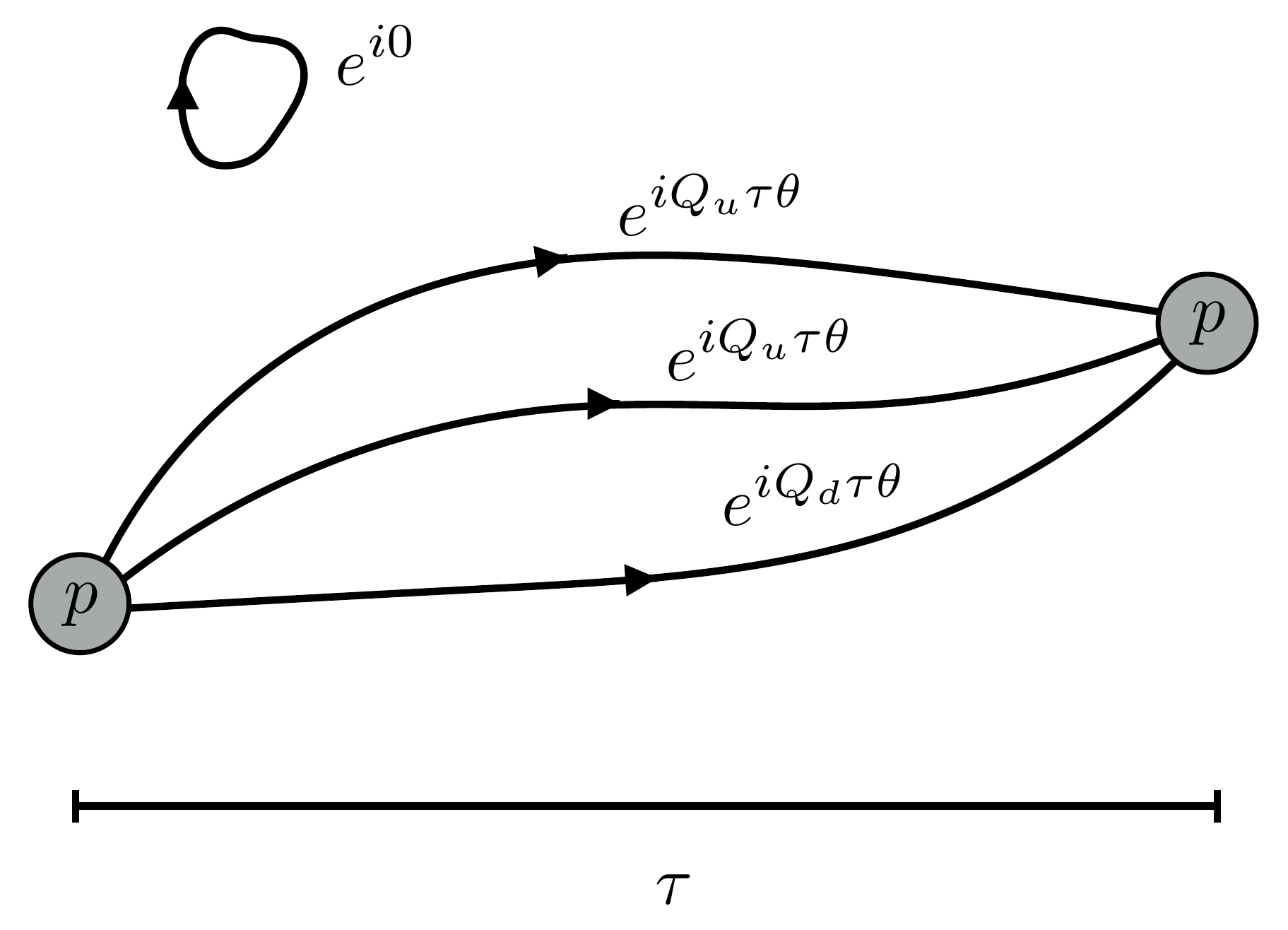} 
\caption{\label{fig:zero_mode_corr}% 
A typical diagram contributing to a two-point correlation function for the proton (two up and one down valence quarks with winding numbers $w=0$) in a hopping parameter expansion.
Each quark line is accompanied by a Wilson line from which the temporal photon zero mode contributes an overall phase.
Since all quark diagrams with winding number $w=0$ have an identical zero mode phase, the phase may be factored out and integrated over analytically (nonperturbatively).
}
%\end{figure}
\end{wrapfigure}

Let us begin by understanding the effects of the massive temporal zero mode $\tilde A_0(0)$ on hadronic correlation functions.
Rather than go into the details of such an accounting from the standpoint of an EFT, we instead provide a heuristic explanation of how temporal zero mode corrections appear.
These effects can be treated nonperturbatively by noting that the temporal zero mode appears precisely in the same manor as an imaginary chemical potential.
From the standpoint of a hopping parameter expansion for a charged two-point function with sources and sinks separated temporally by a distance $\tau$, as shown schematically in \Fig{zero_mode_corr}, diagrams involving quark ($q=u,d,s$) world lines connecting the source and sink are accompanied by a path-independent phase factor $e^{i Q_q \theta (\tau + w T)}$, where $w$ is a temporal winding number and $\theta\equiv \tilde A_0(0)/(L^3 T)$.
Fermion ``bubbles'', however, are independent of $\theta$ since the phases associated with forward and backward propagation in these loops exactly cancel.
These observations allow us to isolate an overall phase factor associated with the leading low-energy contributions to correlation functions, namely $e^{i Q \theta \tau}$, where $Q = \sum_q Q_q$ is a sum over valence quark charges $Q_q$.
Higher winding number contributions are associated with thermal effects, and are suppressed exponentially by comparison to the leading contribution mentioned above.
Noting that the gauge action for the temporal zero mode is given by $S^\theta_\gamma = m_\gamma^2 L^3 T \theta^2 /(8\pi \alpha)$, averaging correlators over the temporal zero mode gives rise to a quadratic dependence on the time separation, given by
\begin{eqnarray}
C(\tau) \propto e^{- M \tau} \int d\theta e^{- S^\theta_\gamma + i Q \theta \tau} \propto e^{- M \tau - x \tau^2}\ ,
\end{eqnarray}
where $x = 2\pi\alpha Q^2/(m_\gamma^2 L^3 T)$ and $0 < \tau \ll T$.
Note that although the quadratic separation dependence is negligible when $2\pi \alpha \ll m_\gamma^2 M L^3$, its size can be quite significant compared to the mass splittings $\Delta M$ which we are interested in, and therefore these corrections must be accounted for in our mass extractions.

\begin{wrapfigure}{R}{0.5\textwidth}
%\begin{figure}
\centering
\includegraphics[width=\figwidth]{\figdir 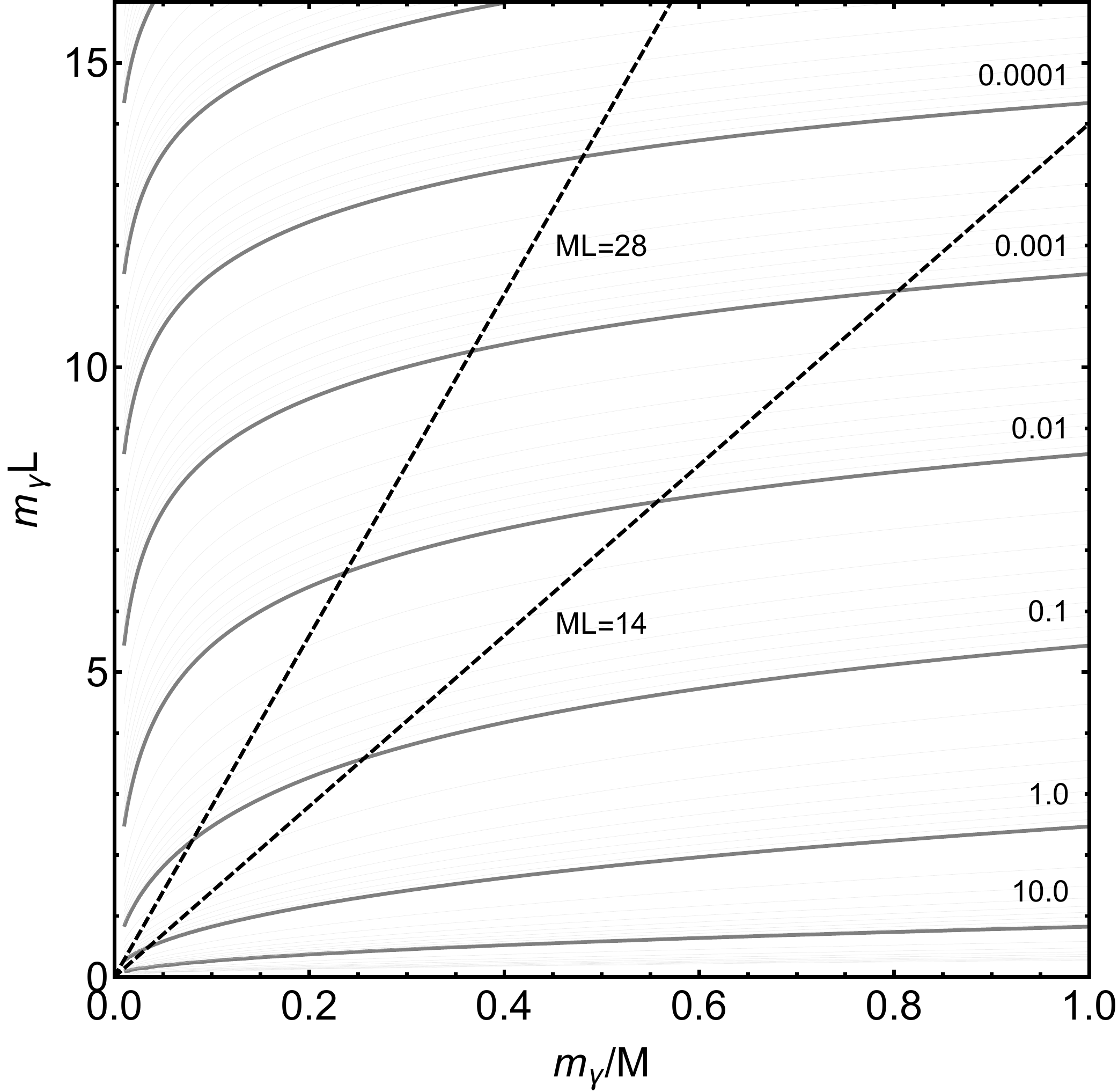} 
\caption{\label{fig:NLOonLOfv}% 
Contour plot of $\delta_L M^{NLO}/\delta_L M^{LO}$ as a function of $m_\gamma/M$ and $m_\gamma L$.
Dashed lines correspond to lines of constant $M L$, along which the $m_\gamma/m_\pi \to 0$ limit is taken (with $m_\pi/M$ held constant).
For our smallest volume studies, $m_\pi L \approx 14$ and $m_p L \approx 28$.
Note that this extrapolation at fixed $ML$ is only reliable provided that the uncorrected finite volume effects are smaller that statistical uncertainties on the mass splittings.
This regime can be identified approximately by the convergence properties of the FV expansion, which are in turn characterized by the contours.
}
%\end{figure}
\end{wrapfigure}

Contributions to mass shifts {\it not associated with the temporal zero mode} are determined to leading order in $\alpha$ using a generalization of nonrelativistic QED for hadrons \cite{Caswell:1985ui}, which appropriately accounts for the gauge symmetry violation of the photon mass term in Landau gauge.
The infinite volume shift in hadron masses due to a finite photon mass are given by $\delta M(\alpha, m_\gamma) \equiv M(\alpha,m_\gamma) - M(\alpha,0)$.
The leading order (LO) and next-to-leading order (NLO) contributions to these shifts are given by
\begin{eqnarray}
\delta M^{LO} &=& -\frac{\alpha}{2} Q^2 m_\gamma \cr
\delta M^{NLO} &=& \left(C e^2 - \frac{\alpha}{4\pi} Q^2 \right) \frac{m^2_\gamma}{M}\ ,
\end{eqnarray}
with the NNLO correction appearing at ${\textrm O}(m_\gamma^3/M^2)$.
Finite volume corrections to hadron masses, on the other hand, are given by $\delta_L M(\alpha,m_\gamma, L) \equiv M(\alpha,m_\gamma, L) - M(\alpha,m_\gamma, \infty)$.
The LO and NLO contributions to the formulas are given by
\begin{eqnarray}
\delta_L M^{LO} &=& 2 \pi \alpha Q^2 m_\gamma\left[ {\cal I}_{1} (m_\gamma L) - \frac{1}{(m_\gamma L)^3} \right] \cr
\delta_L M^{NLO} &=& \pi \alpha Q^2 \frac{m^2_\gamma}{M} \left[ 2 {\cal I}_{1/2}(m_\gamma L) + {\cal I}_{3/2}(m_\gamma L ) \right]\ ,
\end{eqnarray}
where
\begin{eqnarray}
{\cal I}_n(z) = \frac{1}{2^{n+1/2} \pi^{3/2} \Gamma(n)} \sum_{{\bf \nu} \neq {\bf 0}} \frac{K_{3/2-n}(z |{\bf \nu}|)}{(z |{\bf \nu}|)^{3/2-n}}\ ,
\end{eqnarray}
and ${\bf \nu} \in {\mathbb Z}^3$.
To gain a sense of the relative importance of these terms for the parameter regime considered, we provide a contour plot of the ratio $\delta_L M^{LO} / \delta_L M^{NLO}$ as a function of $m_\gamma L$ and $m_\gamma /M$ in \Fig{NLOonLOfv}.
Note that extrapolations in $m_\gamma/m_\pi$ are taken along lines of constant $M L$, assuming $m_\pi/M$ is constant.
Thus, from this plot, we may gauge the level of control over finite volume corrections attained at NLO as a function of our extrapolation parameter $m_\gamma/m_\pi$.

With an analytic understanding of the finite volume and photon mass corrections at hand, we now turn to simulations.
Electroquenched QED+QCD simulations were performed using the Chroma Software System for lattice QCD \cite{Edwards:2004sx}.
Studies were performed using dynamical $SU(3)$ flavor-symmetric QCD gauge ensembles generated on isotropic lattices using a tadpole-improved L{\"u}scher-Weisz gauge action and clover fermion action (see \cite{Beane:2012vq} for further details).
Three ensembles of size 956, 515, and 342 where considered, corresponding to a single $a\approx 0.145$ fm lattice spacing and the lattice volumes $L\approx 3.48$ fm ($L/a=24$), $L\approx 4.64$ fm ($L/a=32$) and $L\approx 6.96$ fm ($L/a=48$), respectively.
All volumes correspond to a pion and kaon mass of $m_\pi = m_K \approx 800$ MeV and nucleon mass of $m_p = m_n \approx 1.6$ GeV.

\begin{figure}[t]
\centering
\includegraphics[width=\textwidth]{\figdir 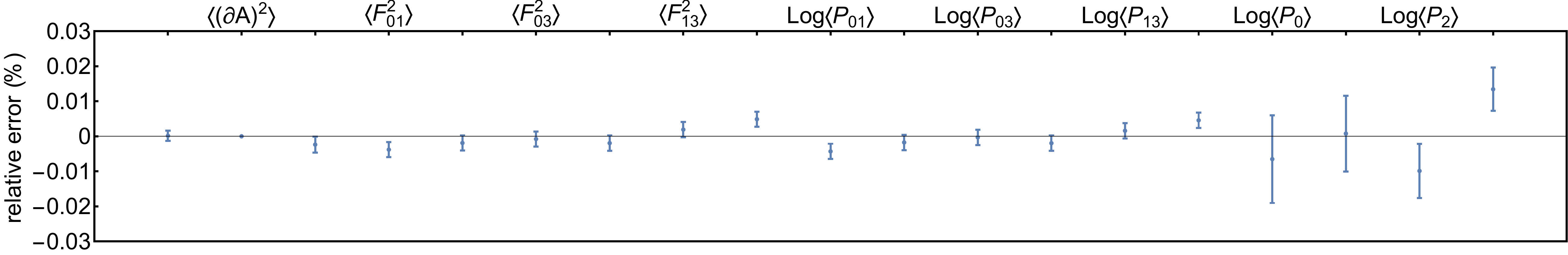} \\
\includegraphics[width=\textwidth]{\figdir 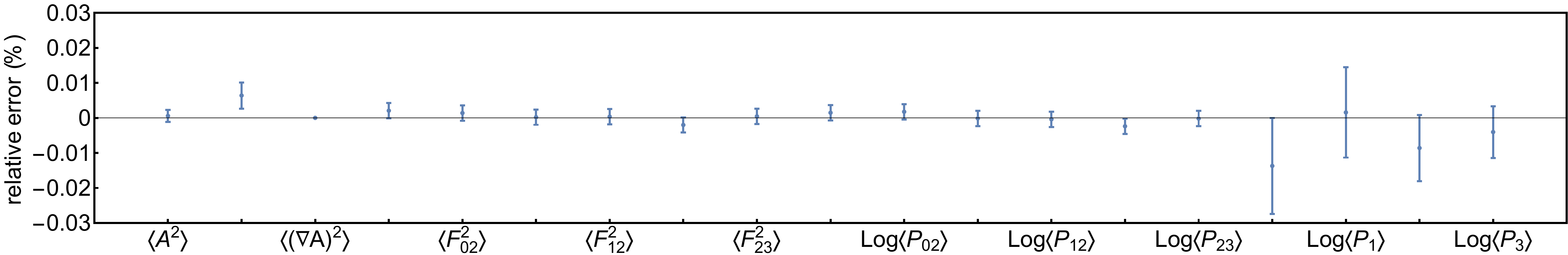}
\caption{\label{fig:u1_test}% 
Comparison of stochastically estimated observables with exact results for massive Landau gauge-fixed (upper) and massless Coulomb gauge-fixed (lower) pure noncompact $U(1)$ gauge theory.
}
\end{figure}

Two sets of $U(1)$ gauge ensembles were generated using a noncompact formalism; the first corresponds to a Coulomb gauge fixed QED action with removal of the zero mode, and the second correspond to the Landau gauge-fixed QED action with photons of mass $m_\gamma/m_\pi = 1/14$, $1/7$, $1/4$, $1/3$, $5/12$, $1/2$, $7/12$ and $1$.
The $U(1)$ gauge configurations for each case were drawn from a Gaussian distribution according to the probability measure ${\cal P}(A) = e^{-S_\gamma(A)}$, with $S_\gamma(A)$ defined in \cite{Portelli:2010yn} for the former case, and defined by \Eq{massive_qed_action} in the latter case.
We performed high precision numerical tests of both our massless Coulomb and massive Landau gauge-fixed codes by comparing numerical estimates of a wide variety of space-time averaged observables with exact analytic calculations.
These observables include functions of the gauge field, and its three- and four-divergences ($\nabla$ and $\partial$, respectively), noncompact and compact definitions of the field strength tensor ($F_{\mu\nu}$ and $P_{\mu\nu} = 2 \cos F_{\mu\nu}-2$), and Polyakov loops ($P_\mu$).
The comparisons were made on a lattice of size $3\times 5\times 7 \times 11$, using a photon mass $a m_\gamma = \gamma \approx 0.57722$.
Results of such comparisons are provided in \Fig{u1_test} for ensembles of size 1M.

Lattice QCD+QED configurations were obtained by post-multiplication of QCD gauge fields by $e^{i e Q_q A_\mu}$ for valence quarks $q$ of charge $Q_q$ ($Q_u=2/3$ and $Q_d = Q_s = -1/3$).
Hadronic correlation functions were subsequently calculated on background QCD+QED configurations in order to determine the mass splittings due to electromagnetic effects.
In the presence of QED, the valence quark masses were tuned such that the $q\bar q$-meson mass $m_{qq}$ extracted from the {\it connected} part of the $q\bar q$ correlation function agreed with $m_\pi$ and $m_K$ at $\alpha=0$.
To that end, all massless and massive QED studies were performed using the valence quark masses $a m_u = -0.25501$ and $am_d=am_s=-0.24750$; by comparison, the QCD bare-quark mass was $am_q=-0.2450$ for all $q$.
Effects of mistuning of the mass splittings were estimated in chiral perturbation theory and found to be on the order of 10\% for the kaon and 25\% for the nucleon.
Given that these mistuning effects are common to both the massless and massive QED studies, we expect such effects to be irrelevant for the purposed of making a comparison of shifts and splittings between the two QED formulations.

Mass shifts and mass splittings were determined for all three lattice volumes in both the cases of QCD with massless and massive QED by studying the late-time dependence of ratios of correlation functions.
In the case of massive QED, differences were determined for all $m_\gamma/m_\pi$, and care was taken to properly account for the quadratic temporal dependence of charged correlators induced by the temporal zero mode.
Details of the interpolating operators used for each correlator, noise reduction techniques, and the extraction of these differences can be found in \cite{Endres:2015gda}.

\begin{figure}
\centering
\includegraphics[width=\figwidth]{\figdir 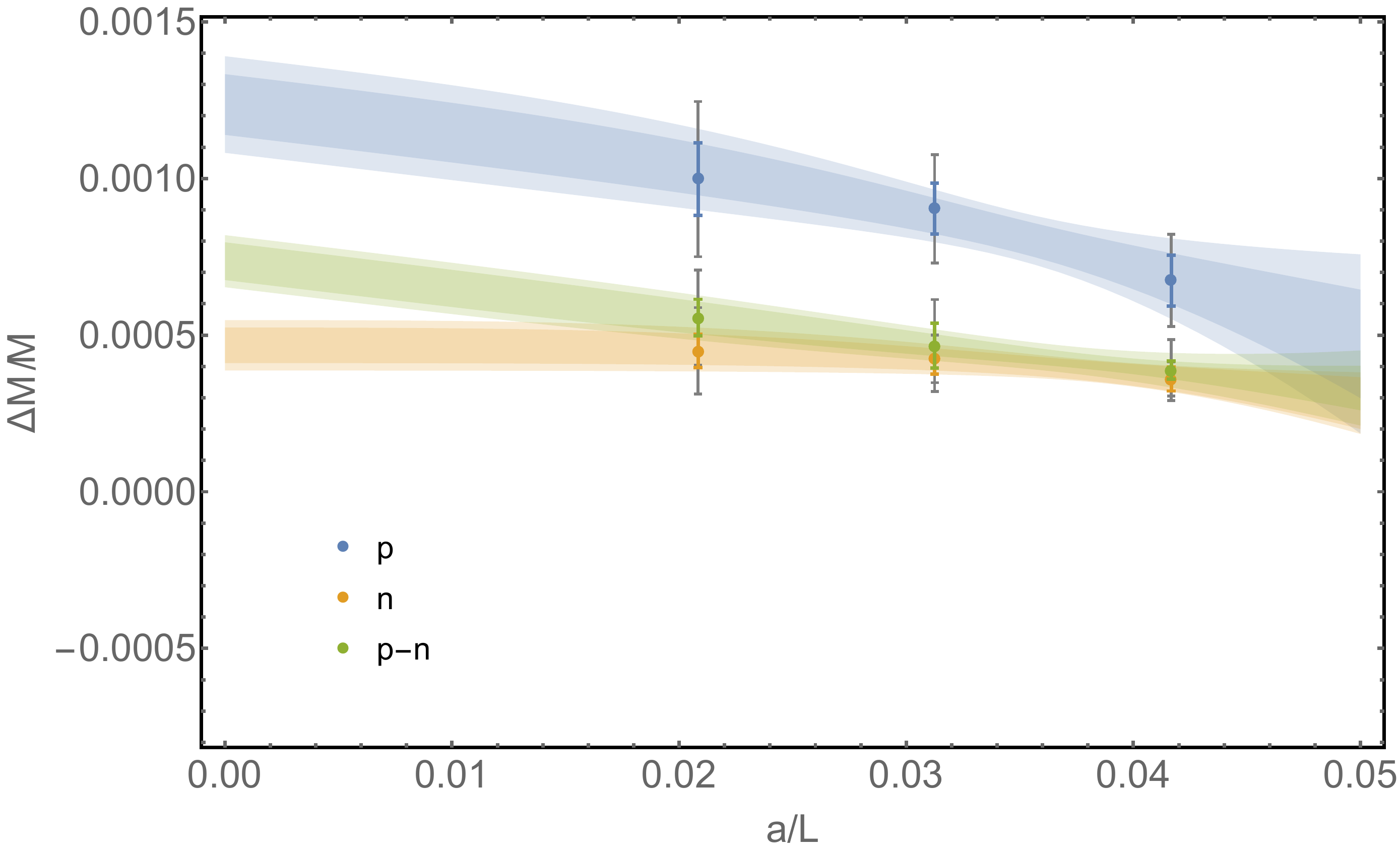}
\includegraphics[width=\figwidth]{\figdir 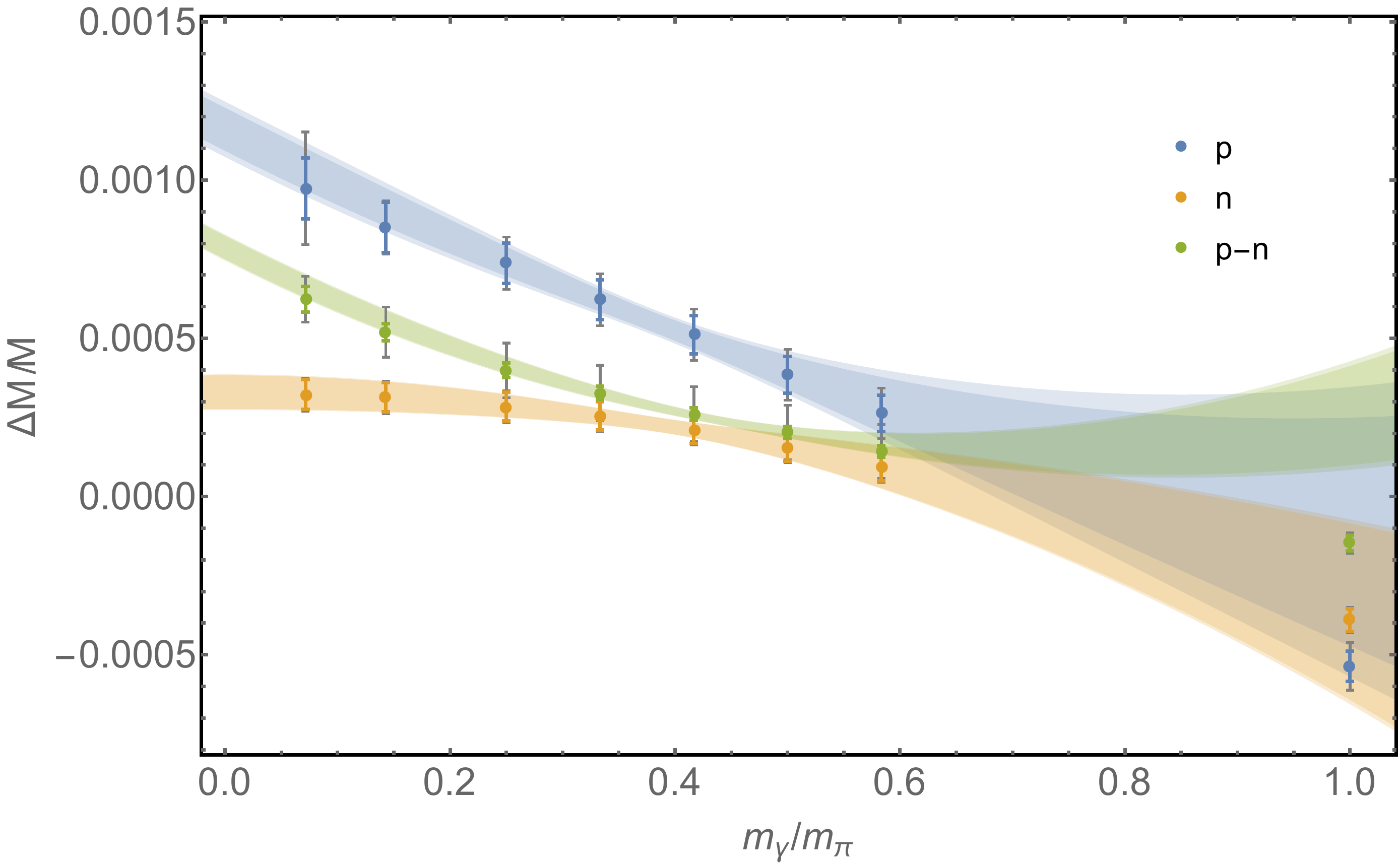} 
\caption{\label{fig:nucleon_diffs}% 
Left: Infinite volume extrapolation of the nucleon mass differences for massless QED.
Right: Vanishing photon mass extrapolation of the nucleon mass differences for massive QED.
}
\end{figure}

Having determined the hadronic mass differences at multiple volumes and multiple photon masses, we utilized appropriate mass shift formulas to perform extrapolations to infinite volume and/or vanishing photon mass.
For the sake of brevity, the discussion of these extrapolations is limited here to several examples involving the nucleons; a more thorough analysis, taking into account all known sources of systematic errors (including variation of fit ranges and order of the fit functions) can be found in \cite{Endres:2015gda}.
In the benchmark case of massless QCD, infinite volume extrapolations of the nucleon  mass differences were performed using the fit formula
\begin{eqnarray}
\Delta M(\alpha,L) = \Delta M(\alpha) + \frac{\alpha Q^2 c_1}{2L} \left(1 + \frac{2a}{L} \right) + \frac{d a^2}{L^3}\ ,
\end{eqnarray}
where $c_1=-2.83729\cdots$, and $\Delta M(\alpha)$ and $d$ are fit parameters.
A representative plot of these extrapolations using all volume data are shown in \Fig{nucleon_diffs} (a), and yield the results $\Delta m_p = 0.00123(10)(14)$, $\Delta m_n = 0.00047(6)(6) $ and $\Delta m_{p-n} = 0.00074(6)(6) $.
In the massive QED case, we use the analytic expressions for $\delta_L M$ to remove the LO and NLO contributions to volume dependence in the extracted mass differences determined at each volume.
Note that this constitutes a fit-less extrapolation of the data to infinite volume, up to systematic errors associated with volume dependence at NNLO, which is not removed; the size of these effects can be qualitatively assessed from the convergence of the expansion illustrated in \Fig{NLOonLOfv}.
For example, for $m_\gamma/m_p \gtrsim 1/4$ ($m_\gamma/m_\pi \gtrsim 1/8$), one finds $\delta_L M^{NLO}/\delta_L M^{LO} < 0.1$ for the nucleon.
We subsequently, perform $m_\gamma/m_\pi\to0$ extrapolations in the volume corrected data, using the fit function
\begin{eqnarray}
\Delta M(\alpha,m_\gamma) = \Delta M(\alpha) + \delta M^{LO}(\alpha,m_\gamma) + \delta M^{NLO}(\alpha,m_\gamma) \ ,
\end{eqnarray}
where $\Delta M(\alpha)$ and $C$ (implicit in $\delta M^{NLO}$) are fit parameters.
Examples of such fits for mass splittings determined from our smallest $L = 3.48$ fm ensembles for $m_\gamma/m_\pi \in [1/4,1/2]$ are provided in \Fig{nucleon_diffs} (b) and yield the results $\Delta m_p = 0.00116(7)(5)$, $\Delta m_n = 0.00033(5)(2)$ and $\Delta m_{p-n} = 0.00079(4)(2)$.

From a full analysis of the data at all volumes discussed in \cite{Endres:2015gda}, we find that our $m_\gamma/m_\pi$ extrapolations are generally robust against variation in fit ranges and extrapolation order, and as demonstrated in the illustrative example above, yields results for mass splittings which are consistent with conventional infinite volume extrapolations in the massless QED formulation.
We find these results to be true not only for the nucleon mass shifts and splittings, but also for the kaon, which was not discussed here.
The results, suggest that massive QED is a viable alternative to conventional methods; the advantages of using this approach over conventional methods are discussed in greater detail elsewhere \cite{Endres:2015gda}.

\begin{acknowledgments}
We would like to thank W.~Detmold, R.~Edwards, B.~Jo\'{o}, K.~Orginos and D.~Richards for the JLab/W\&M QCD gauge field configurations used in this work.
Additionally, we would like to thank D.~B.~ Kaplan, T.~C.~Luu, and M.~J.~Savage for useful conversations and correspondences and A.~Patella and N.~Tantalo for stimulating discussions during the Lattice 2015 conference.
We would like to acknowledge the hospitality of the International Institute of Physics at the Federal University of Rio Grande de Norte and the Institute for Nuclear Theory at the University of Washington (Nuclear Reactions Workshop~\cite{Briceno:2014tqa}), where portions of this work were completed.
Computations for this study were carried out on facilities of the USQCD Collaboration, which are funded by the Office of Science of the U.S. Department of Energy.
M. G. E was supported by the the U. S. Department of Energy Early Career Research Award DE-SC0010495, and moneys from the Dean of Science Office at MIT.
B. C. T. was supported in part by a joint City College of New York-RIKEN/Brookhaven Research Center fellowship, a grant from the Professional Staff Congress of the CUNY, and by the U.S. National Science Foundation, under Grant No. PHY15-15738.
A. W-L. was supported in part by the U.S. Department of Energy (DOE) contract DE-AC05-06OR23177, under which Jefferson Science Associates, LLC, manages and operates the Jefferson Lab and by the U.S. DOE Early Career Award contract DE-SC0012180.

\end{acknowledgments}

\bibliography{massive_qed}
\bibliographystyle{h-physrev}

\end{document}